\documentclass[onecolumn,10pt,nofootinbib,preprintnumbers]{revtex4-2}
\usepackage[linewidth=1pt]{mdframed}
\usepackage[margin=0.6in]{geometry}
\usepackage[utf8]{inputenc}
\usepackage{amsmath}
\usepackage{graphicx}
\usepackage{physics,amssymb}
\usepackage{hyperref}
\hypersetup{
}

\usepackage{makecell}
\newcolumntype{?}{!{\vrule width 0.1pt}}
\def\as{\alpha_s}
\def\asb{\bar{\alpha}_s}

\def\cW{\mathcal{W}}

\def\cC{\mathcal{C}}
\def\cF{\mathcal{F}}
\def\cG{\mathcal{G}}
\def\cO{\mathcal{O}}

\def\R{\footnotesize R}
\def\CF{\mathrm{C_F}}
\def\CA{\mathrm{C_A}}

\def\d{\text{d}}
\def\inn{\mathrm{in}}

\begin{document}

\title{Jet shapes in H/V boson + jet with $k_t$ clustering at hadron colliders}
\author{K. Khelifa-Kerfa}\email{kamel.khelifakerfa@univ-relizane.dz (Speaker)}
\affiliation{Physics Department, Faculty of Science and Technology,\\  Universit\'{e} de Relizane - Relizane 48000, Algeria}
\author{Y. Delenda}\email{yazid.delenda@univ-batna.dz}
\author{N. Ziani}\email{naima.ziani@univ-batna.dz}
\affiliation{Laboratoire de Physique des Rayonnements et de leurs Intéractions avec la Matière, Département de Physique, Faculté des Sciences de la Matière,\\ Université de Batna-1, Batna 05000, Algeria}

\begin{abstract}
We present analytical calculations of the distribution of non-global jet shapes in Higgs/vector boson + jet production at hadron colliders. Within the eikonal-limit framework and implementing various jet algorithms, we compute the full distribution of the particular jet mass shape observable at 2-loops, including the large single-logarithms known as non-global logs and clustering logs. We compare our next-to-leading-log analytical resummation to parton showers and, after matching and including non-perturbative effects, to experimental data. A good agreement is shown for all comparisons.

\bigskip
{\it Presented at DIS2022: XXIX International Workshop on Deep-Inelastic Scattering and Related Subjects, 
May 2 - 6, 2022, Santiago de Compostela, Spain.}
\end{abstract}

\maketitle

\flushbottom

\section{Introduction}

The study of jet shapes and jet substructure have recently seen huge attention due to their effectiveness in investigating both (background) QCD and new physics (signals) (see ref. \cite{Banfi:2004yd} for a full review). The invariant mass of a jet, in particular, plays a key role in scrutinising various aspects of QCD including, to mention a few, initial- and final-state radiation, colour flow, soft and/or collinear regions, hadronisation, and underlying event. The analytical calculations of this jet shape, and others, are complementary to Monte Carlo (MC) simulations as they address many issues that are vaguely clear in the latter. These include, for instance, theoretical uncertainties and quantification of missing higher-terms. 

The jet mass is a member of a family of observables known as "non-global" observables (NGO) \cite{Dasgupta:2001sh, Dasgupta:2002bw}. Unlike ``global" observables, NGO are sensitive only to particular regions of the phase space (instead of the full phase space). In addition to the usual soft/collinear large logarithms that arise in the perturbative distribution of jet (and event) observables, NGO are plagued with other large logarithms known as ``non-global logs" (NGLs). Moreover, when jets are clustered using jet algorithms \cite{Catani:1993hr, Ellis:1993tq, Dokshitzer:1997in, Wobisch:1998wt, Cacciari:2008gp} NGO receive another ladder of large logarithms known as "clustering logs" (CLs). A proper next-to-leading-log (NLL) resummation of the jet mass distribution has to account for all three types of logs mentioned above. It is this very task that we address in what follows below. 
 
The resummation of the jet mass observable is performed for the particular process of the production of a single hard jet in association with a vector boson ($W, Z$ or $\gamma$) or a Higgs boson $H$ at the Large Hadron Collider (LHC). Previous works on the said observable and within the Soft and Collinear Effective Theory (SCET) include: di-jet events \cite{Liu:2014oog}, $\gamma$+jet events \cite{Chien:2012ur} and H+jet events \cite{Jouttenus:2013hs}. The current work extends that of Ref. \cite{Dasgupta:2012hg} from various aspects including: 
\begin{itemize}
\item various jet algorithms, namely: anti-k$_t$, k$_t$ and Cambridge-Aachen (C-A). 
\item various processes, namely: Z/W/$\gamma$/H + jet. 
\item full jet-radius dependence of both NGLs and CLs at two-loops order. 
\item compare results of all four processes (NLL resummation + next-to-leading-order (NLO) matching) with various parton showers.
\item estimating hadronisation and underlying event corrections for the Z+jet process and comparing the full result with CMS experimental data reported in Ref. \cite{Chatrchyan:2013rla}. 
\end{itemize}    

In the next section we present the details of the kinematics of the processes and the observable to be considered herein.

\section{Setup}

The four hadronic processes, i.e., $W/Z/\gamma/H$+jet are identical from the point of view of QCD calculations as they all involve three hard coloured QCD partons and a colour-neutral boson $X$. They only differ in the corresponding Born partonic channels: $(\delta_1): q \bar{q} \to g + X, (\delta_2): q g \to q + X$ and $(\delta_3): gg \to g + X$. 
In all of our calculations we shall assume eikonal (or equivalently soft) approximation with strong ordering in transverse momenta. The said assumptions significantly simplify the calculations while retaining the necessary NLL accuracy. Recoil effects are also ignored as they are beyond NLL accuracy. All partons are considered massless throughout. 

The normalised (squared) invariant mass observable of a jet $j$ is defined by:
\begin{align}\label{eq:JetMassDef}
\varrho = \frac{1}{p_t^2} \left[ p_j + \sum_{i \in j} k_{i} \right]^2  = \sum_{i\in j} \varrho_i, \quad
\varrho_i = \frac{2 (k_i \cdot p_j)}{p_t^2},
\end{align}
where $p_j$ and $p_t$ are the four- and transverse momentum of the outgoing measured jet, $k_i$ is the four-momentum of the $i^{\mathrm{th}}$ soft emission and the sum is over all emissions that end up inside the jet after the application of the jet clustering algorithm. Three jet algorithms are considered in the present work, namely, anti-k$_t$, k$_t$ and C-A, as mentioned in the introduction. 

The (differential) distribution of the jet mass for a given Born channel $\delta$ is given by: 
\begin{align}\label{eq:dSigma}
 \frac{\d \Sigma_\delta(\rho)}{\d B_\delta} = \int_0^\rho \frac{\d \sigma_\delta}{\d B_\delta \d \varrho} \d \varrho =  \frac{\d \sigma_{0,\delta}}{\d B_\delta} \, f_{B, \delta} (\rho) C_{B, \delta}(\rho),
\end{align}
where $\d\sigma_{0,\delta}/\d B_\delta$ is the differential partonic Born cross-section for channel $\delta$ with respect to the Born configuration $B_\delta$, $C_{B, \delta}(\rho) = 1 + \as\, C^{(1)}_{B, \delta} + \cdots $ is a function with non-logarithmically-enhanced terms, and $f_{B, \delta}(\rho)$ is a function that resums all  large logarithms
\begin{align}
 f_{B, \delta}(\rho) = \exp\left[ L g_1(\as L) + g_2(\as L) + \cdots   \right],
\end{align}
where $L g_1$ resums leading (double) logs that originate from soft and collinear emissions off the hard parton initiating the jet, and $g_2$ resums NLL (single) logs that come from various sources, including: (a) hard-collinear emissions from outgoing hard partons, (b) soft wide-angle emissions from all partons, (c) NGLs from correlated soft wide-angle secondary emissions, and (d) CLs from soft wide-angle primary emissions when jet algorithms other than anti-k$_t$ are used. 

Before presenting the NLL resummed form factor for the jet mass observable we first show the fixed-order calculations for one- and two-gluons emissions.

\section{Fixed-order calculations}

\subsection{One-gluon emission}

The three Born channels mentioned above with a soft emission $k_1$ may be schematically represented as 
\begin{align}
a + b \to j + X + k_1.
\end{align}
The jet mass perturbative distribution at one-loop may be cast in the form:
\begin{align}\label{eq:f1}
 f^{(1)}_{B, \delta} (\rho) = - \int \d\Phi_1 \, \cW_{1, \delta}^{\R} \, \Theta(\varrho_1 - \rho) \, \Xi_\inn (k_1),
\end{align}
where the expressions the phase space factor $\d\Phi_1$ and the one-loop eikonal amplitude squared (for real emission) $\cW_{1, \delta}^{\R}$ are given in our paper \cite{Ziani:2021dxr}. The clustering function $ \Xi_\inn (k_1)$ restricts the soft emission $k_1$ to be inside the jet for it to contribute to its mass. The final result reads 
\begin{align}
 f^{(1)}_{B, \delta} (\rho) = - (\cC_{aj} + \cC_{bj}) \asb \frac{L^2}{4}  - \asb L \left[ \cC_{ab}\, \frac{R^2}{2} + (\cC_{aj} + \cC_{bj}) h(R) \right], 
\end{align} 
where $\cC_{ij}$ is the colour factor associated with the dipole ($ij$): $\cC_{q \bar{q}} = 2 \CF-\CA = -1/N_c, \cC_{qg} = \cC_{gg} = \CA = N_c$ with $\CF$ and $N_c$ having their usual meaning, $L = \ln(R^2/\rho)$ is the large log to be resummed, and $h(R) = R^2/8 + R^4/576 + \cO(R^8)$.

\subsection{Two-gluon emissions}

The eikonal squared amplitude at two-loops order consists of two parts: a {\it reducible} part which corresponds to primary emissions off the three-hard-legs Born configuration and accounts for global Abelian logarithms and CLs if k$_t$ and C-A are used; and an {\it irreducible} part which corresponds to secondary non-Abelian emissions and accounts for NGLs. The jet mass distribution at two-loops may be written as 
\begin{align}
 f^{(2)}_{B, \delta}(\rho) = \frac{1}{2!} \left(f^{(1)}_{B, \delta} \right)^2 + C_{2, \delta} (\rho) + S_{2, \delta} (\rho), 
\end{align}
where $C_{2, \delta} (\rho)$ accounts for CLs and $S_{2, \delta} (\rho)$ accounts for NGLs. 

Unlike pure global Abelian logarithms that are resummed by the famous Sudakov form factor, CLs are not captured by the said form factor and need to be computed at each perturbative order. At two-loops they are given by
\begin{align}
 C_{2, \delta}(\rho) = \frac{1}{2!} \asb^2 L^2\, \cF_2^\delta(R).
\end{align}
The expressions of $\cF_2^\delta(R)$ for the various Born channels are given in our paper \cite{Ziani:2021dxr}. Recall that for anti-k$_t$ CLs are absent. 

NGLs, on the other hand, are given at two-loops, by 
\begin{align}
 S_{2, \delta}(\rho) = \frac{1}{2!} \asb^2 L^2\, \cG_2^\delta(R).
\end{align}
The expressions of $\cG_2^\delta(R)$ for all three jet algorithms and all Born channels are given in our paper \cite{Ziani:2021dxr}. As is well established by now, the effect of jet clustering algorithms, other than anti-k$_t$, is twofold: on one hand, they reduce the size of NGLs, and on the other hand, they introduce CLs into the distribution. 

\begin{figure}[h!]
\centering
\includegraphics[scale=0.5]{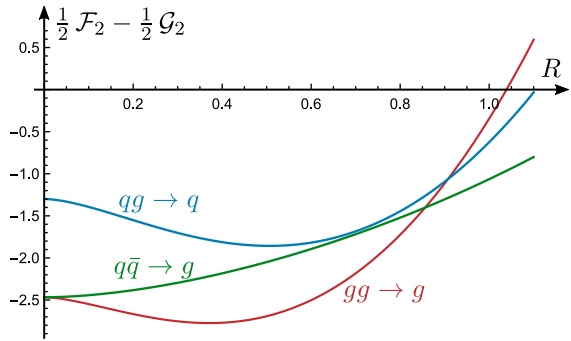}
\caption{Combined effect of CLs and NGLs at two-loops with k$_t$ clustering.} \label{fig:F2-G2}
\end{figure}

Figure \ref{fig:F2-G2} shows the overall coefficient of CLs and NGLs in the jet mass distribution at two-loops. The two large logs tend to balance each other out for large jet radii ($R \gtrsim 1$).
We remark that for small values of $R$ one recovers results found in $e^+ e^- \to$ di-jet events \cite{KhelifaKerfa:2011zu}.

\section{Resummation and comparison to parton showers}

\begin{figure}[h!]
\centering
\includegraphics[scale=0.45]{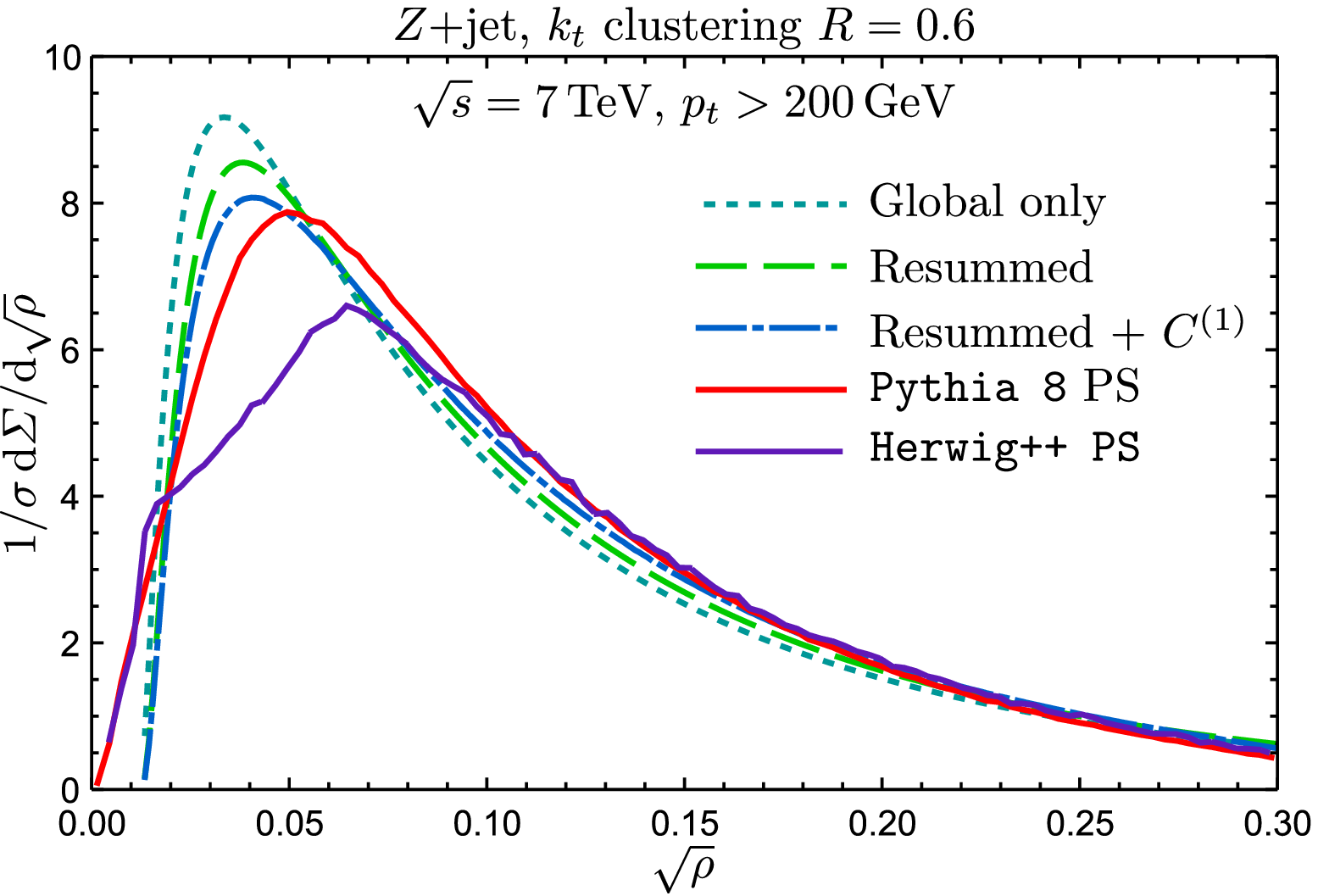}\\
\includegraphics[scale=0.45]{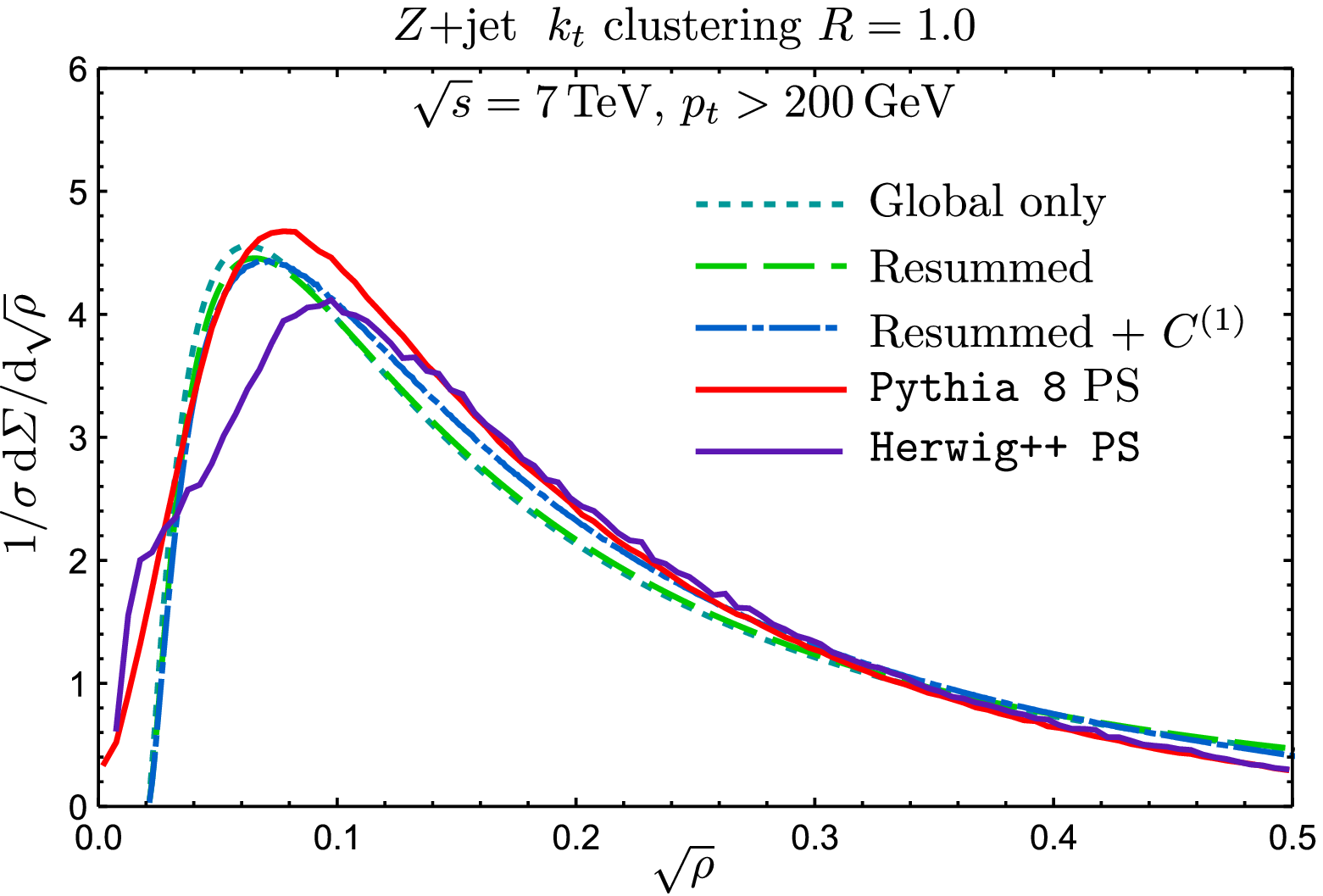}
\includegraphics[scale=0.45]{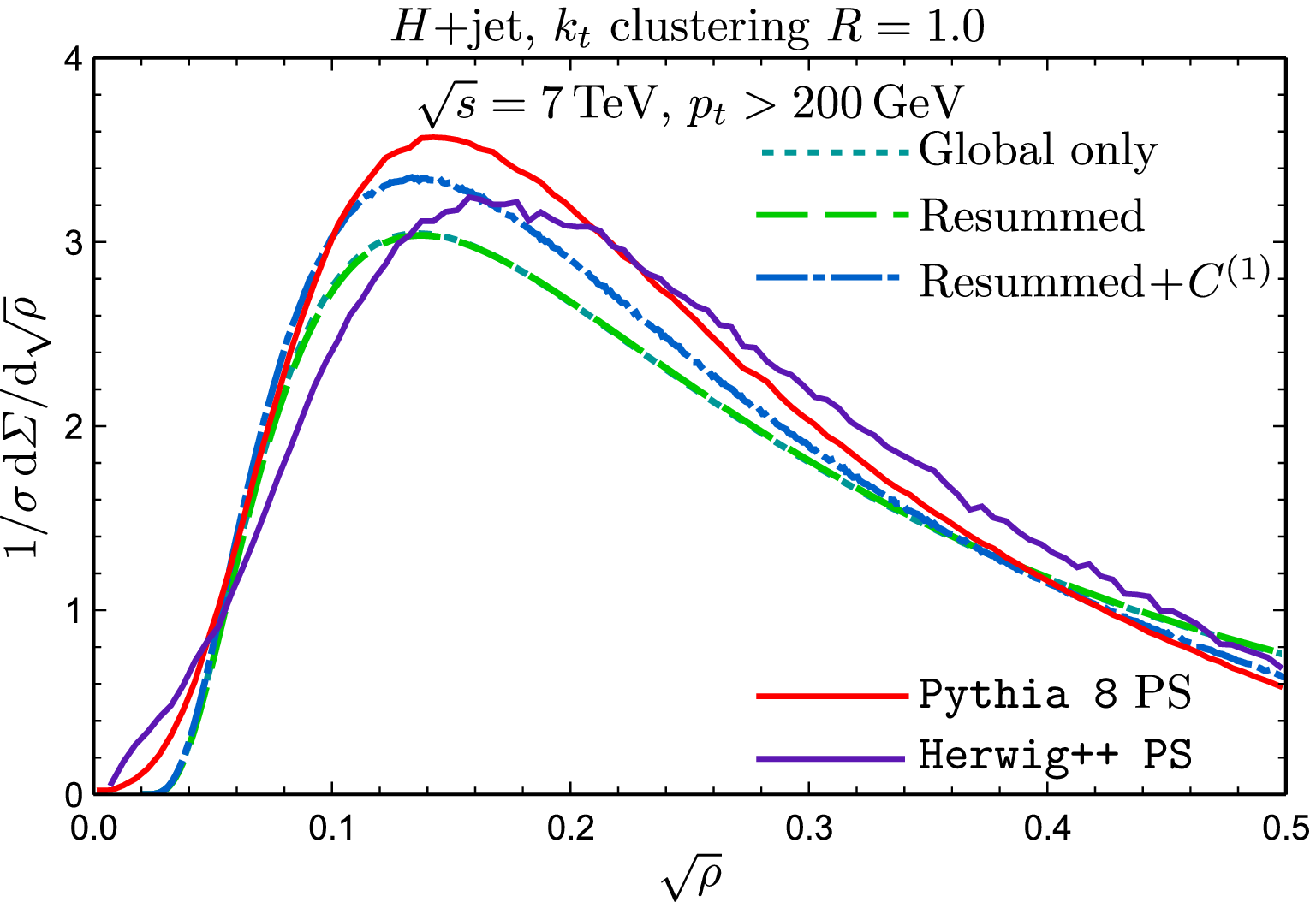}\\
\includegraphics[scale=0.38]{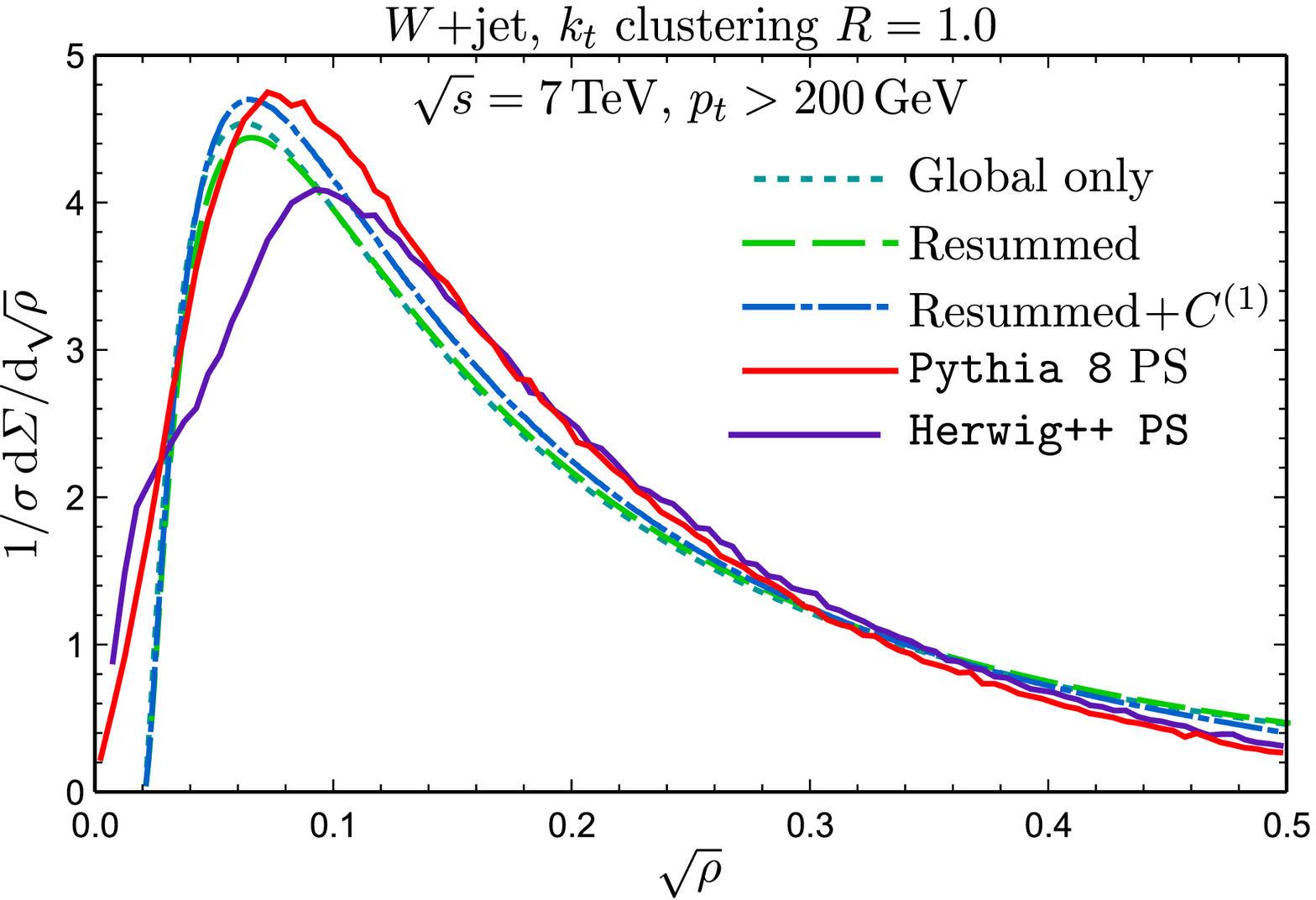}
\includegraphics[scale=0.38]{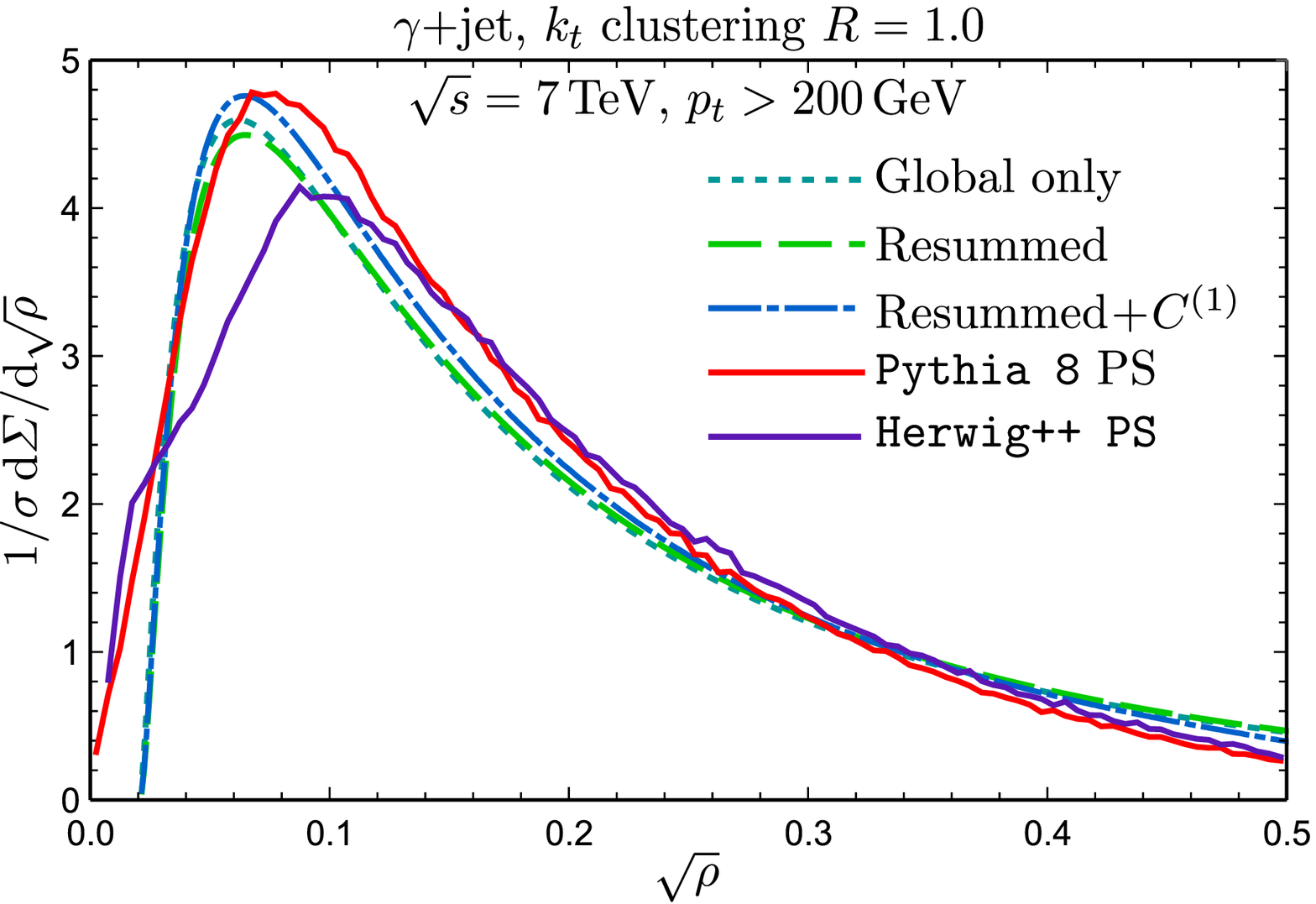}
\caption{Comparisons of the resummed formula \eqref{eq:NLL-resum-formula} for  the various processes to parton showers.} \label{fig:Comparison}
\end{figure}

The NLL-resummed jet mass distribution is given, after including NGLs and CLs, by:
\begin{align}\label{eq:NLL-resum-formula}
 \frac{\d \Sigma_\delta(\rho)}{\d B_\delta} = \frac{\d \sigma_{0,\delta}}{\d B_\delta} \, f^{\mathrm{global}}_{B, \delta} (\rho)\, C_\delta(\rho) \, S_\delta(\rho)\, C_{B, \delta}(\rho),
\end{align}
where $C_\delta(\rho)$ and $S_\delta(\rho)$ resum CLs and NGLs, respectively, and are approximated by the exponential of the two-loop result. They may be determined to all-orders only numerically both at large- and finite-N$_c$ for a few observables (not including our jet mass observable) \cite{Hagiwara:2015bia, Hatta:2020wre}. An estimate of the NLO effect on the distribution is included in eq. \eqref{eq:NLL-resum-formula} via the term $C^{(1)}_{B, \delta}$ obtained from the fixed-order program \texttt{MCFM} \cite{Campbell:2015qma}. Figure \ref{fig:Comparison} presents comparisons of the NLL-resummed formula \eqref{eq:NLL-resum-formula} to various parton showers for the processes Z/H/W/$\gamma$ + jet. 
There is generally a good agreement of the resummed + $C^{(1)}$ distribution with \texttt{Pythia 8} parton shower for all four processes and for the values of $R$ considered. NGLs and CLs are sizeable for $R = 0.6$ compared to $R = 1.0$, as was hinted at in the previous section. Moreover, the inclusion of the one-loop constant term $C^{(1)}_{B,\delta}$ seems to improve the distribution both at the peak and tail regions. 
We note that discrepancies between the two parton showers may be lifted off once non-perturbative effects are included, which we consider in the next section. 

\section{Comparison to CMS data}

\begin{figure}[h!]
\centering
\includegraphics[scale=0.58]{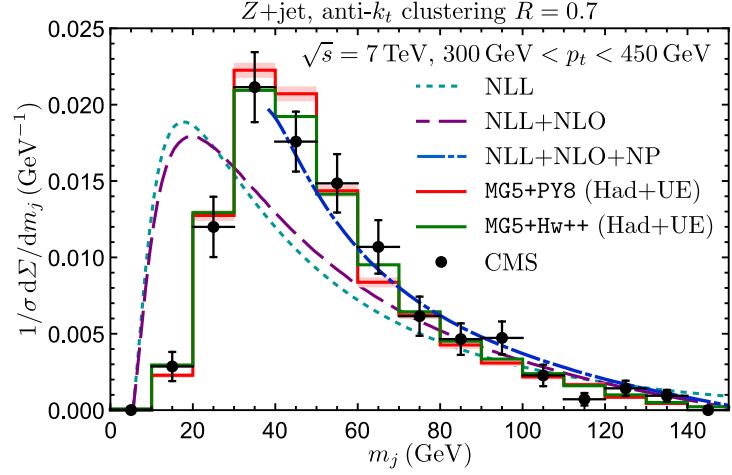}
\caption{Comparison of analytical calculations, parton showers and CMS experimental data for the (un-normalised) jet mass variable $m_j$.} \label{fig:CMS}
\end{figure}

To compare to experimental data from CMS \cite{Chatrchyan:2013rla, 1224539/t32} we first match our NLL-resummed formula \eqref{eq:NLL-resum-formula} to NLO exact distribution taken from \texttt{MCFM}. We then analytically estimate the size of the two dominating non-perturbative (NP) effects, namely, hadronisaion and underlying-event using the method of the shift in the mean value of the jet mass \cite{Dasgupta:2007wa}. Figure \ref{fig:CMS} shows the comparison of the NLL-resummed and matched formula with associated NP corrections with CMS data and results from \texttt{MadGraph}~5 \cite{Maltoni:2002qb} interfaced to \texttt{Pythia}~8 \cite{Alwall:2008qv} and \texttt{Herwig}~7 for Z+jet in anti-k$_t$ with $R = 0.7$ and jet $p_t$ in the range $300 < p_t < 450$ GeV. 
A good agreement is shown between resummed prediction, CMS data and event generators over a wide range of the observable. The cut-off at around $40$ GeV of the NLL+NLO+NP distribution is a manifestation of the shift method, which renders the distribution valid only to the right of the peak. 

\section{Summary}

In this proceeding, we have presented analytical calculations of an important jet-shape observable, namely, the invariant mass of a jet, both at fixed-order and to all-orders in perturbation theory. The latter observable being of non-global nature poses delicate challenges to resum. We have shown how one can overcome such challenges and presented an NLL resummation that includes all-orders estimate of NGLs and CLs. Comparisons to various parton showers as well as to experimental data confirm our calculations for a wide range of values of the observable considered. As future work, it is worth investigating other jet-shapes and other hadronic processes that constitute important backgrounds to numerous new physics signals. 


\bibliography{Refs}

\end{document}